\documentclass[conference]{IEEEtran}
\usepackage{cite}
\usepackage{amsmath,amssymb,amsfonts}
\usepackage{algorithmic}
\usepackage{graphicx}
\usepackage{textcomp}
\usepackage{xcolor}
\usepackage{booktabs}
\usepackage{stfloats}
\usepackage{amsmath, nccmath}
\usepackage{empheq}
\usepackage[subnum]{cases}
\usepackage{comment}
\usepackage{lipsum}
\usepackage{graphicx}

\usepackage{fancyhdr}

\IEEEoverridecommandlockouts

\newcommand*{\rom}[1]{\expandafter\@slowromancap\romannumeral #1@}
\newcommand{\RomanNumeralCaps}[1]
    {\MakeUppercase{\romannumeral #1}}

\usepackage[
singlelinecheck=false 
]{subcaption}
\usepackage{subcaption}
\def\BibTeX{{\rm B\kern-.05em{\sc i\kern-.025em b}\kern-.08em
    T\kern-.1667em\lower.7ex\hbox{E}\kern-.125emX}}
\begin{document}

\title{A novel dual-stream time-frequency contrastive pretext tasks framework for sleep stage classification \\

}

\author{
    \IEEEauthorblockN{
        Sergio Kazatzidis, Siamak Mehrkanoon\IEEEauthorrefmark{1}\thanks{*Corresponding author.} 
    }
    \IEEEauthorblockA{Department of Information and Computing Sciences, Utrecht University, Utrecht, Netherlands}
    s.kazatzidis@gmail.com, 
    s.mehrkanoon@uu.nl
}

\maketitle
\begin{abstract}
Self-supervised learning addresses the challenge encountered by many supervised methods, i.e. the requirement of large amounts of annotated data. This challenge is particularly pronounced in fields such as the electroencephalography (EEG) research domain. Self-supervised learning operates instead by utilizing pseudo-labels, which are generated by pretext tasks, to obtain a rich and meaningful data representation. In this study, we aim at introducing a dual-stream pretext task architecture that operates both in the time and frequency domains. In particular, we have examined the incorporation of the novel Frequency Similarity (FS) pretext task into two existing pretext tasks, Relative Positioning (RP) and Temporal Shuffling (TS). We assess the accuracy of these models using the Physionet Challenge 2018 (PC18) dataset in the context of the downstream task sleep stage classification. The inclusion of FS resulted in a notable improvement in downstream task accuracy, with a 1.28 percent improvement on RP and a 2.02 percent improvement on TS. Furthermore, when visualizing the learned embeddings using Uniform Manifold Approximation and Projection (UMAP), distinct clusters emerge, indicating that the learned representations carry meaningful information.
\end{abstract}

\begin{IEEEkeywords}
Self-supervised learning, Sleep Staging, Pretext tasks
\end{IEEEkeywords}

\section{Introduction}
Large annotated data sets are generally necessary to train deep neural networks. However, obtaining these data sets can be time-consuming and costly \cite{Jing2020}. Self-supervised learning has recently garnered considerable attention due to its capacity to circumvent the need for labeled data, addressing this fundamental challenge. This paradigm utilizes self-generated pseudo-labels during the pre-training phase. These pseudo-labels are generated by pretext tasks in order to obtain a meaningful representation of the data. The choice of the appropriate pretext tasks is important in capturing the general features of the data. Furthermore, these learned features could be utilized in downstream tasks such as classification and segmentation \cite{Jais2020}. Self-supervised learning has been successfully applied in diverse fields such as Natural Language Processing \cite{Bert2018,Word2vec2017,GPT3-2020}, and Computer Vision \cite{simclr2020,Jigsaw2016,Moco2020}. However, the utilization of Self-supervised learning in the field of EEG processing is still in its early phases. This is notable given the resource-intensive nature of obtaining annotations in EEG research \cite{SSLYiang2021}. For instance, annotating a 24-hour EEG sleep signal segment requires roughly five hours of concentrated effort by an expert \cite{sleepscoring}.

The significance of sleep quality for overall health highlights the need for thorough investigation, including the classification of sleep stages, which can aid in detecting sleep disorders like sleep apnea. Deep learning networks, particularly Convolutional Neural Networks (CNNs), offer a viable approach for this classification task  \cite{sleepdeeplearn2020} and various other tasks \cite{CNN2021,siamak2019}. However, this method still requires a substantial volume of labeled data. The authors in \cite{Banville2021} devised three pretext tasks, Relative Positioning (RP), Temporal Shuffling (TS), and Contrastive Predictive Coding (CPC), to facilitate the learning of meaningful representations of the data without the requirement for large volumes of labeled data. In this paper, we specifically focus on RP and TS. RP is a pretext task where the model determines whether a second segment of EEG-data is temporally close or distant from the first segment, which is referred to as the anchor segment, operating on the assumption that nearby EEG data is more similar, while more distant data is less similar. In TS the model determines whether three windows of EEG-data are ordered or shuffled in time. In this paper we present an improvement of the aforementioned pretext tasks, by incorporating the novel Frequency Similarity (FS) task. To this end, we propose a dual-stream pretext tasks architecture that concatenates the learned features of RP or TS with those of FS, introducing a frequency component to the features to enhance the data representation. In the FS pretext task the model evaluates whether a window exhibits greater similarity in the frequency domain to the anchor window than another window.

This paper is organized as follows. Section \RomanNumeralCaps{2} provides a brief overview of related work, while Section \RomanNumeralCaps{3} covers preliminaries. Section \RomanNumeralCaps{4} introduces the FS pretext task and the novel architecture. Section \RomanNumeralCaps{5} showcases the obtained results, which are subsequently discussed. The conclusion is drawn in Section \RomanNumeralCaps{6}.  

\section{Related works}

\subsection{Self-supervised learning}

The major challenge of supervised machine learning techniques is their heavy reliance on the amount and the quality of the labeled data. Data annotation can be a laborious and expensive endeavor, motivating the exploration of self-supervised learning as a viable alternative. In self-supervised learning one uses the unannotated data to learn the underlying structure and to obtain a rich representation of the data. The adoption of the right pretext tasks is often pivotal in learning meaningful data representations \cite{Jais2020}. Self-supervised learning has been used mainly in the field of Computer Vision \cite{simclr2020,Jigsaw2016,Moco2020} and Natural Language Processing \cite{Bert2018,Word2vec2017,GPT3-2020}, but in other fields such as Audio Processing \cite{audio2020,audio2022}, Video Processing \cite{video2016,video2021,video2023} and EEG Processing \cite{EEG2022,EEG2022-2} self-supervised learning is rapidly gaining popularity. Self-supervised learning approaches can be broadly categorized into generative and contrastive methods \cite{Jais2020}. Generative methods aim to reconstruct the original input while learning a meaningful representation \cite{survey2021}. In contrast, contrastive methods adopt a discriminative approach, seeking to group similar samples closely while pushing diverse samples apart \cite{Jais2020}.
 An example of the generative method can be observed in Bert \cite{Bert2018}, where certain tokens within a sentence are masked and predicted by the model. An early instance of the contrastive method in Computer Vision involves utilizing a jigsaw puzzle to learn the data's structure by reassembling shuffled patches back into the original image \cite{Jigsaw2016}.

In this paper, we will focus on contrastive learning, a method that clusters similar samples together while increasing the distance between dissimilar data pairs. Positive samples, signifying similarity, are created through various data augmentation techniques such as random cropping, resizing, color distortions, and rotations \cite{Jais2020}. These techniques are applied to the original data to generate positive data pairs. Any other data is considered negative and dissimilar with respect to the original data.
A successful framework in contrastive learning that employs these data augmentations is SimCLR \cite{simclr2020}, which uses two different data augmentation techniques on the original image and maximizes their agreement in order to find the underlying structure of the original image. Another prominent contrastive framework is Momentum Contrast (MoCo) \cite{Moco2020}, which introduces momentum updates to stabilize the rate of representation change, enhancing representation consistency.  

\subsection{EEG-signals}
This central focus of this paper revolves around EEG data, with the primary objective of uncovering its underlying structures and patterns for sleep stage classification. In EEG research, data annotation is a laborious and costly process, primarily carried out by domain experts. Moreover, the noise in the data can complicate the visual interpretation. Self-supervised learning offers a promising solution to these challenges by utilizing pseudo-labels generated through pretext tasks \cite{Banville2021}. Various self-supervised learning approaches have been employed to unveil the underlying patterns in EEG data. For instance, in studies like \cite{EEG2021} and \cite{EEG2020} multiple data augmentations are applied to EEG data, including techniques such as time warping, Gaussian noise, additive noise, amplitude scale, average filter and more. These augmentations are used to create positive pairs from the same data. The aim is to bring their representations close together, while increasing the gap between these positive pairs and other negative pairs, which consist of the original data and any other dissimilar data. Another approach, as described in \cite{EEG2022-3},  involves devising a pretext task for the network to solve. In this case, a task called \textquotedblleft Temporal Rearrange'' is introduced, where a window of EEG data is divided into three blocks and shuffled (or not). The model's task is to predict whether shuffling occurred, thereby learning meaningful representations from the EEG data. In some cases, multiple pretext tasks are learned together, a strategy known as multi-task learning. For instance, in a recent study \cite{EEG2022-2}, the authors simultaneously trained three distinct pretext tasks: one focusing on spatial aspects, another centered around frequency-related features, and the third utilizing contrastive learning. In contrast, here we focus on learning features separately through the time-based and frequency-based pretext tasks and subsequently concatenating them. In particular, we utilize Relative Positioning (RP) and Temporal Shuffling (TS) proposed in \cite{Banville2021} for the time-based pretext tasks. We will then combine the learned features of RP or TS with features learned by a novel frequency-based task that will be introduced in Section 4. After learning and extracting the valuable and necessary information from these pretext tasks, the knowledge is applied to a downstream task. An example of
a downstream task where the information from the pretext tasks can be utilized is sleep stage
classification. In this task the model must classify the EEG data into one of the five
stages of sleep \cite{SSLYiang2021}, which is crucial for identifying sleep disorders. Currently, sleep stage scoring in EEG data is predominantly performed manually by experts, a time-consuming and error-prone process \cite{sleepdeeplearn2020}. The proposed self-supervised learning methods in this paper hold the potential to automate and improve this process.

\section{Preliminaries}
\label{sec:sample1}

In this section the two utilized pretext tasks, i.e. Relative Positioning and Temporal Shuffling \cite{Banville2021}, will be briefly explained.
\subsection{Relative Positioning}
The Relative Positioning task (RP) requires a pair of two windows. The underlying assumption of this pretext task is that neighboring windows correspond to similar sleep stages, while distant windows represent different sleep stages. The first window, referred to as the anchor window, is denoted by $x_t$. The second window, denoted by $x_{t'}$, is selected based on two key parameters $\tau_{pos}$ and $\tau_{neg}$. The parameter $\tau_{pos}$ indicates the time frame around the anchor window that is considered close by. Positive pairs are composed of the anchor window and another window within this positive context. The parameter $\tau_{neg}$ defines the negative context. This parameter indicates the minimum time frame from the anchor window that is deemed sufficiently distant. Windows falling inside this negative context are eligible to be utilized as negative samples, forming negative pairs with the anchor window. The index $t$ associated with each window indicates their starting time, i.e. the window $x_{t'}$ has starting time $t'$ . The labeling of the $i$-th pair is then determined as follows \cite{Banville2021}:
\begin{equation}
y_i = \begin{cases}
0,\   |t_{i}-t'_i| > \tau_{neg} 
\\
or
\\
1,\   |t_{i}-t'_i| \leq \tau_{pos}.
\end{cases}
\end{equation}
In the study of RP, here in this paper a window of 30 seconds is utilized. In addition, the parameters $\tau_{pos}$ and $\tau_{neg}$ are set to 1 minute and 15 minutes respectively.
\begin{figure*}[]

\begin{subfigure}{\textwidth}
\includegraphics[width=1.0\linewidth]{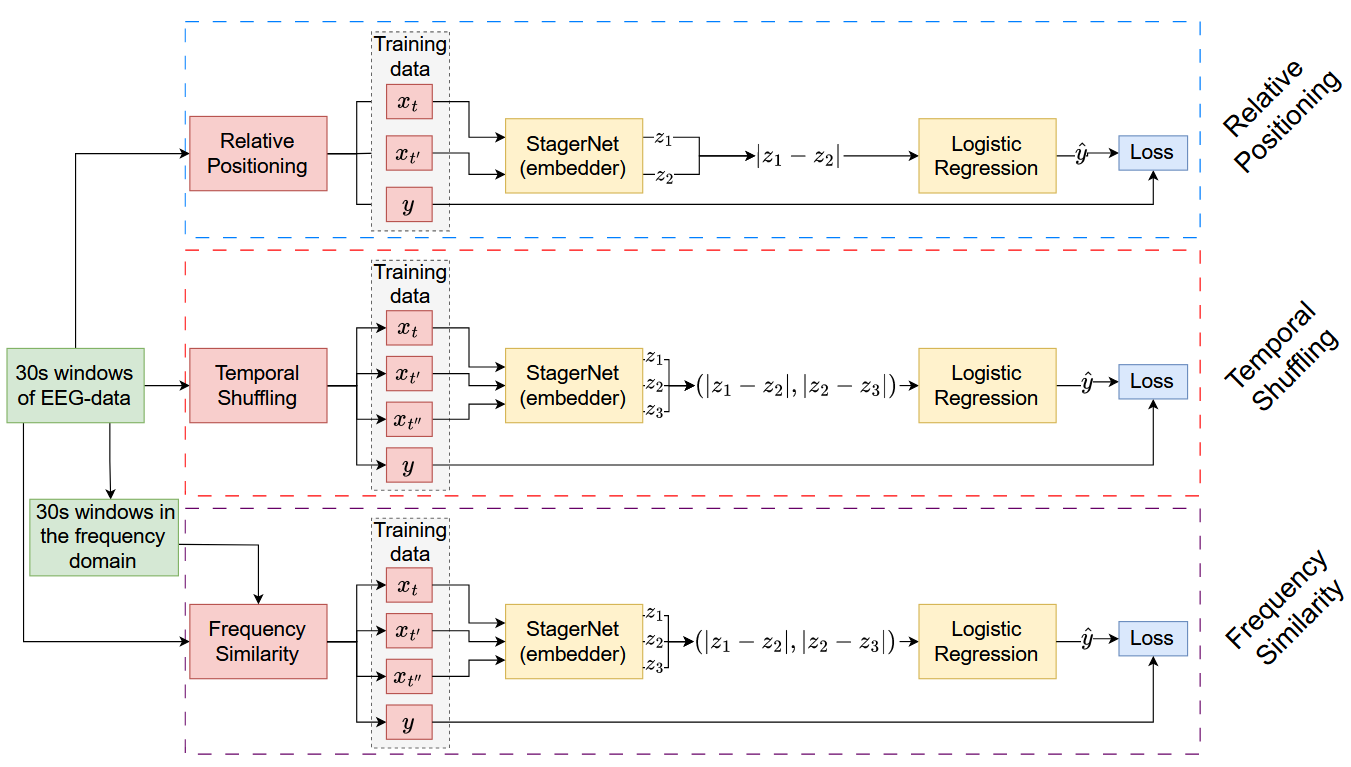}

\caption{}
\label{fig:prea}
\end{subfigure}

\begin{subfigure}{\textwidth}
\centering
\includegraphics[width=0.85\linewidth,]{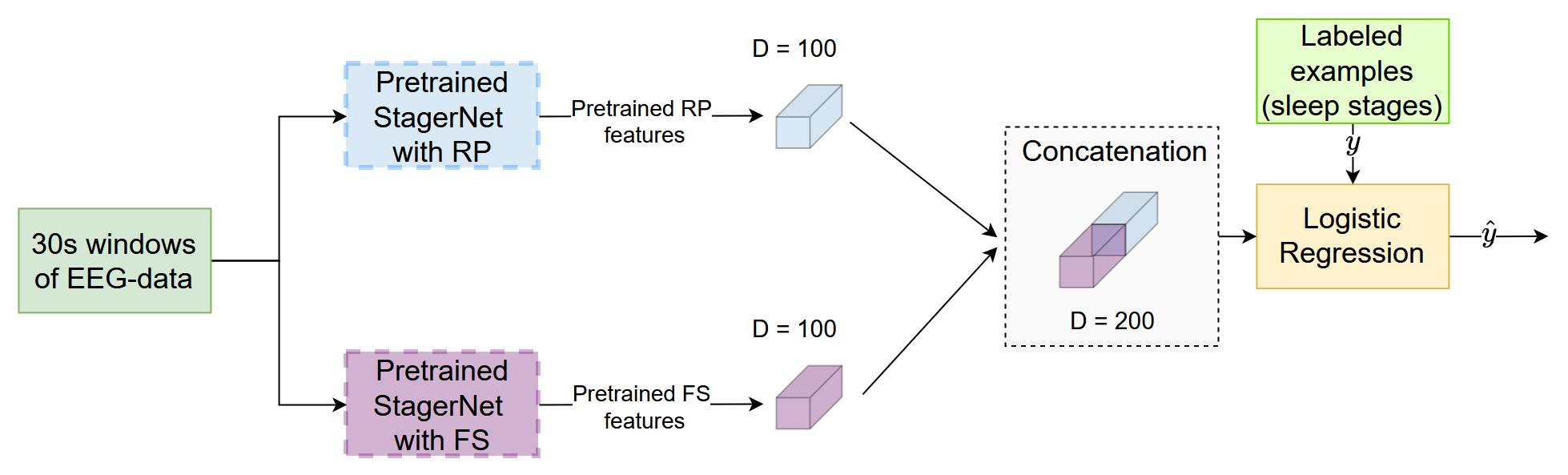} 
\caption{}
\label{fig:subdownb}
\end{subfigure}
\begin{subfigure}{\textwidth}
\centering
\includegraphics[width=0.85\linewidth,]{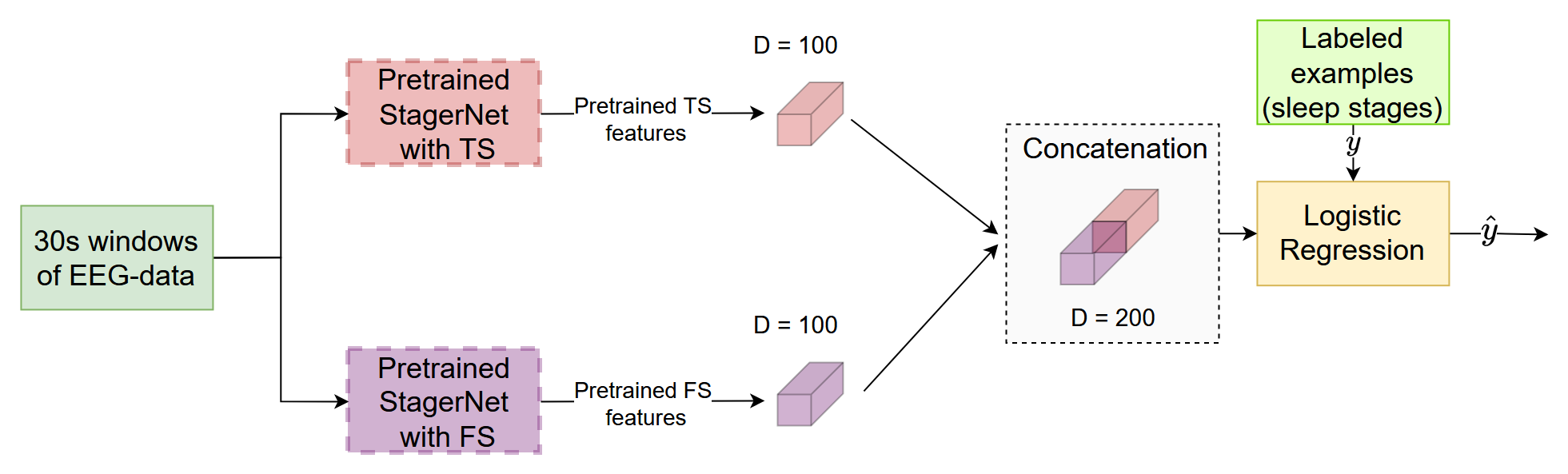} 
\caption{}
\label{fig:subdownc}
\end{subfigure}
\caption{(a) Architectural framework for pretext task learning. (b) Dual-stream time-frequency framework with RP and FS for the downstream task. (c) Dual-stream time-frequency framework with TS and FS for the downstream task.}
\label{fig:multitask}
\end{figure*}
To learn from this task two operations are defined: $h_{\theta}$ and $g_{RP}$. $h_{\theta}$ is a feature extractor (embedder), in our case StagerNet \cite{stager2018}, which maps a window to a representation in the feature space. $g_{RP}$ is a contrastive module that combines feature embeddings using element-wise absolute differences. An overview of the RP process is shown in Fig. \ref{fig:prea} within the blue dashed box. Ultimately, a linear model with a dropout rate of 50\% is employed to predict whether a given pair is positive or negative.

\subsection{Temporal Shuffling}
The Temporal Shuffling (TS) pretext task requires three windows, denoted as $x_t$, $x_{t'}$, $x_{t''}$.  Initially, we randomly select the first window, denoted as the anchor window $x_t$. Subsequently, another window $x_{t''}$ is chosen to be in close temporal proximity to the anchor window, guided by the parameter $\tau_{pos}$. The third window, $x_{t'}$, can be positioned either in between windows $x_{t}$ and $x_{t''}$ or in the negative context based on $\tau_{neg}$. The parameter $\tau_{neg}$ indicates the minimum time frame from the anchor window that is deemed sufficiently distant. Following this, we proceed to create window triplets that can take one of two forms: they are either arranged in temporal order ($t < t' < t''$) or shuffled ($t < t'' < t'$ or $t' < t < t''$), with $t$ representing the starting time of window $x_t$. The labeling of the $i$-th triplet is then determined as follows \cite{Banville2021}: 

\begin{equation}
y_i = \begin{cases}
0, \   t'_i 	\notin \ [t_{i},t''_i]\ \
\\
or
\\
1, \   t'_i 	\in \ ]t_{i},t''_i[\;.\ \
\end{cases}
\end{equation}

In the study of TS, here in this paper a window of 30 seconds is utilized. In addition, the parameters $\tau_{pos}$ and $\tau_{neg}$ are set to 30 minutes and 120 minutes respectively. The TS learning process closely mirrors that of Relative Positioning, and is visually represented within the red dashed box in Fig. \ref{fig:prea}. After obtaining the embeddings $h_\theta(x_t)$, $h_\theta(x_{t'})$ and $h_\theta(x_{t''})$, the contrastive module $g_{TS}$ concatenates the absolute differences, $|h_\theta(x_t)-h_\theta(x_{t'})|$ and $|h_\theta(x_{t'})-h_\theta(x_{t''})|$, as follows \cite{Banville2021}:

\begin{equation}
\begin{IEEEeqnarraybox}[][c]{rCl}
g_{TS}(h_\theta(x_t),h_\theta(x_{t'}),h_\theta(x_{t''})) & = & (|h_\theta(x_t)-h_\theta(x_{t'})|, \\ && |h_\theta(x_{t'})-h_\theta(x_{t''})|).
\end{IEEEeqnarraybox}
\label{eq:eq3}
\end{equation}

\section{Method}

\subsection{Frequency Similarity}
We propose a novel pretext task, referred to as  Frequency Similarity (FS), that operates in the frequency domain. Three distinct non-overlapping windows ($x_t$,$x_{t'}$,$x_{t''}$) are randomly sampled from the EEG recording, with $x_t$ serving as the anchor window. The only requirement is that these windows must be distinct from one another. Subsequently, the windows $x_{t}$, $x_{t'}$ and $x_{t''}$ are transformed from the time domain into the frequency domain utilizing the Welch method \cite{welch2014}, resulting in $x_{f}$, $x_{f'}$ and $x_{f''}$. The Welch method is a well-established PSE technique that estimates the distribution of power across different frequency components within an EEG signal. This transformation allows the pretext task to operate in the frequency domain, enabling the assessment of relationships between these frequency representations. The Hellinger Distance (HD) is chosen as the metric to assess the distance between the transformed windows $x_f$ and $x_{f'}$, as well as between $x_f$ and $x_{f''}$. This choice is motivated by a study in \cite{similarity2020}, which demonstrated the effectiveness of the Hellinger Distance for analyzing EEG data in the frequency domain.
The assignment of the label $y_i$ is based on whether the distance from $x_{f''}$ to $x_f$ is less than or greater than the distance from $x_{f'}$ to $x_f$. 

The labeling of the $i$-th triplet then is determined as follows:
\newline
\begin{equation}
\begin{IEEEeqnarraybox}[][c]{rCl}
y_i = \begin{cases}
0, \ \frac{1}{N} \sum_{i=1}^{N} \frac{1}{\sqrt{2}} \cdot \lVert x_{f}^{ch_i}-x_{f'}^{ch_i} \rVert _2 \\ \ \ \ \  <  \frac{1}{N} \sum_{i=1}^{N} \frac{1}{\sqrt{2}} \cdot \lVert x_{f}^{ch_i}-x_{f''}^{ch_i} \rVert _2
\\
or
\\
1, \ \frac{1}{N} \sum_{i=1}^{N} \frac{1}{\sqrt{2}} \cdot \lVert x_{f}^{ch_i}-x_{f'}^{ch_i} \rVert _2 \\ \ \ \ \ >    \frac{1}{N} \sum_{i=1}^{N} \frac{1}{\sqrt{2}} \cdot \lVert x_{f}^{ch_i}-x_{f''}^{ch_i} \rVert _2\;,
\end{cases}
\end{IEEEeqnarraybox}
\end{equation}
where N represents the total number of channels.

\begin{figure*}[h]

\includegraphics[width=\linewidth]{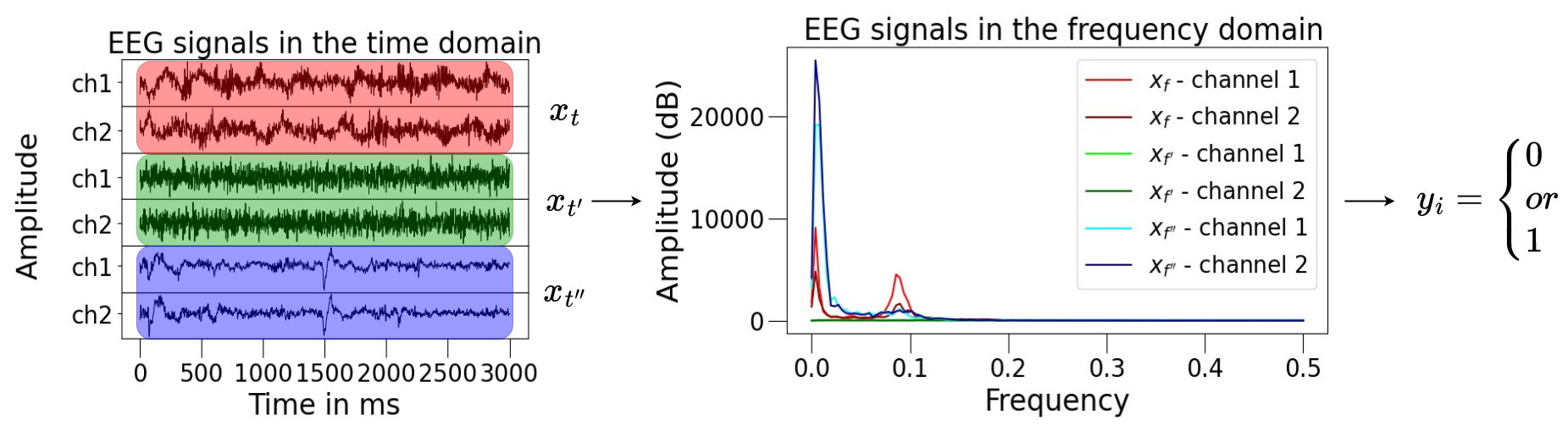}

\caption{Visual explanation of the introduced Frequency Similarity (FS) pretext task. The EEG
inputs are transformed with the Welch method into the frequency domain. Subsequently, the distances between the windows are computed, and labels are assigned based on these distances.
}
\label{fig:SSLFS}
\end{figure*}

The feature extraction process in FS operates similarly to RP and TS. This process is depicted in Fig. \ref{fig:prea} within the dashed purple box. The windows are mapped to a representation in the feature space using $h_{\theta}$, which, in this case, is the StagerNet \cite{stager2018}. It's important to note that the StagerNet receives the three windows $x_t$, $x_{t'}$ and $x_{t''}$ of the time domain as input, but the pseudo-labels are determined in the frequency domain. $g_{FS}$ aggregates the window representations by concatenating the absolute differences between them, similar to $g_{TS}$ as described in Eq. (\ref{eq:eq3}). Subsequently, the output of the $g_{FS}$ is fed into a linear model with weights $w$ and bias weight $w_0$ which is utilized to predict whether the distance from $x_{f''}$ to $x_f$ is less than or greater than the distance from $x_{f'}$ to $x_f$. A dropout rate of 50\% is applied to the linear model layer. The selected loss function is binary cross-entropy, combined with a sigmoid function. This choice is based on the nature of the output of the linear model. The mathematical expression of the loss function is as follows:
\begin{equation}
\mathcal{L}
 = - \left[ y \cdot \log(\sigma(\hat{y})) + (1 - y) \cdot \log(1 - \sigma(\hat{y})) \right],
\end{equation}
where $y$ is the target value, $\hat{y}$ is the prediction of the linear model and $\sigma$ is the sigmoid function. A visualization of the pretext task Frequency Similarity is provided in Fig. \ref{fig:SSLFS}.

\subsection{The proposed dual-stream architecture}
In our approach, each pretext task, i.e. RP, TS and FS, trains its own separate feature extractor denoted as $h_{\theta}$. Following pre-training, these feature extractors utilize the original EEG dataset as input and generate their new feature representation. Subsequently, the features obtained from RP and TS are then individually concatenated with the features of FS, as depicted in Fig. \ref{fig:subdownb} and Fig. \ref{fig:subdownc}. This results in twice as many learned features, encompassing information from both the original time-based pretext tasks and frequency-based FS. As a result, the concatenated features encapsulate knowledge of both time and frequency domains of the EEG data. These combined features are subsequently frozen and utilized in the logistic regression for the downstream task.

\subsection{Downstream task}
The downstream task addressed in this paper is sleep staging, which has been previously studied in the literature, primarily using supervised deep learning methods \cite{sleepCNN2020,sleeptransformer2022}. However, more recent research papers, such as \cite{Banville2021,EEG2021,EEG2020},  have demonstrated the successful application of Self-Supervised Learning (SSL) for this task. Sleep staging involves classifying epochs of sleep EEG-data into one of the five sleep stages: W (awake), N1, N2, N3 (different sleep levels), and R (REM-sleep). The epochs considered here have a duration of 30 seconds. Automating this task is crucial as it currently relies heavily on experts' manual work \cite{sleepstaging2022}. 

\subsection{Stagernet embedder}
The chosen architecture for the embedder, denoted as $h_{\theta}$, is the StagerNet \cite{stager2018}. StagerNet is a convolutional neural network (CNN) designed to process EEG windows as input. It begins by applying a spatial convolution to the input, followed by a permutation and a temporal convolution. Subsequently, it employs a max-pooling layer, followed by another temporal convolution, another max-pooling layer. Finally, it flattens the data and applies dropout, resulting in a feature vector with 100 dimensions. To ensure a fair comparison with the embedder in \cite{Banville2021}, we employed the same version of StagerNet as the embedder. This modified version includes twice the number of convolutional channels (16) compared to the original StagerNet \cite{stager2018} and also includes batch normalization. The StagerNet employs an Adam Optimizer \cite{adamopt2014} with parameters $\beta_1 = 0.9$ and $\beta_2 = 0.999$, and a learning rate $\alpha =$ 5 × $10^{-4}$. The training process was set to run for a maximum of 70 epochs or until the validation loss stopped decreasing for a minimum of 10 consecutive epochs.

\subsection{Data}
The models in this study are trained on a subset of the Physionet Challenge 2018 EEG \cite{datachall2018} dataset, specifically the 994 participants from the training set for whom sleep annotations were available. To create the training, validation, and testing sets, a random split of 60\% for training, 20\% for validation, and 20\% for testing is performed, resulting in 595, 199, and 199 recordings in each respective set. For the pretext tasks RP, TS, and FS, 2,000 pairs or triplets of windows were randomly sampled from each recording. This EEG dataset comprises six EEG channels (F3-M2, F4-M1,C3-M2, C4-M1, O1-M2 and O2-M1) and employs a sampling frequency of 200 Hz. The stages are categorized as W (wakefulness), N1, N2, N3 (non-REM sleep stages) and R (REM sleep). Following the lines of \cite{Banville2021}, first the EEG-waves are filtered applying a 30 Hz Finite Impulse Response lowpass filter with a Hamming window, to filter out higher frequencies, which are not relevant for sleep staging. In addition, as in \cite{Banville2021}, only the channels F3-M2 and F4-M1 are retained to reduce data dimensionality. Subsequently, the data is downsampled to 100 Hz. Non-overlapping 30-second windows are subsequently extracted and then normalized per channel to have zero-mean and unit standard deviation. The proposed code is available at GitHub\footnote{https://github.com/Serg99io/DSTF-sleepstaging}.

\section{Results and discussion}

\label{sec:sample1}
\subsection{Sleep stage classification with different pretext tasks}
In this section, we evaluate the performance of the discussed pretext tasks, Relative Positioning (RP), Temporal Shuffling (TS), and Frequency Similarity (FS), within the context of the downstream task of sleep stage classification. This evaluation is conducted utilizing all available labeled examples, amounting to a total of 521,943 labeled examples.
During the initial phase of pre-training, we train distinct embedders for each of the pretext tasks RP, TS and FS. Subsequently, the training, validation and test sets are provided as input to these pre-trained embedders to obtain their respective representations. Each distinct embedder yields a feature vector with 100 dimensions, capturing the essential information found in an EEG window. The features obtained with FS are then concatenated with those of RP and TS separately, resulting in a feature vector with a dimensionality of 200. For sleep stage classification, a linear logistic regression model with L2-regularization is employed. We have selected weighted average precision, weighted average recall and balanced accuracy as our primary evaluation metrics. The metrics are formally expressed as follows:
\begin{numcases}{}
     Weighted \ avg \ precision = \frac{1}{N}  \sum_{i=1}^{C} n_{c}  \frac{TP_i}{TP_i + FP_i} \ , \tag{6}  \\    
    Weighted \ avg \ recall = \frac{1}{N}  \sum_{i=1}^{C} n_{c}  \frac{TP_i}{TP_i + FN_i} \  ,  \tag{7} \\
    Balanced \ accuracy = \frac{1}{C} \sum_{i=1}^{C} \frac{TP_i}{TP_i + FN_i} \ \  ,  \tag{8}
\end{numcases}
where $N$ represents the total number of instances, $C$ represents the number of classes, $n_{c}$ represents the number of instances per class, $TP_i$, $FP_i$ and $FN_i$ represent the number of true positives, false positives and false negatives for class $i$ respectively. The obtained results are summarized in Table \ref{accuracy}. 
\begin{table}[h]
\centering

\caption{
Balanced accuracy and additional metrics for sleep stage classification on the Physionet Challenge 2018 EEG dataset utilizing all labeled examples.}

\resizebox{0.49\textwidth}{!}{\begin{tabular}{lccccc}
\label{accuracy}

\\ \hline 
  \begin{tabular}[c]{@{}c@{}}\textbf{Pretext}\\ \textbf{tasks}\end{tabular} &
  \begin{tabular}[c]{@{}c@{}}\textbf{Balanced}\\ \textbf{accuracy}\end{tabular} &
  \begin{tabular}[c]{@{}c@{}}\textbf{Weighted avg}\\ \textbf{precision}\end{tabular} &
  \begin{tabular}[c]{@{}c@{}}\textbf{Weighted avg}\\ \textbf{recall}\end{tabular} \\ \hline
 RP \cite{Banville2021}      & 0.6942 & 0.69 & 0.67 \\
 TS \cite{Banville2021}    & 0.6829 & 0.68 & 0.66 \\
 FS      & 0.6365 & 0.65 & 0.62 \\
 RP + FS & \textbf{0.7070} & \textbf{0.70} & \textbf{0.69} \\
 TS + FS & \textbf{0.7031} & \textbf{0.70} & \textbf{0.68} \\ \hline

\end{tabular}}

\end{table}

One can observe that incorporating FS with either RP or TS leads to noticeable improvements in accuracy compared to using RP or TS individually.

\begin{figure}[h!]

\begin{subfigure}{0.44\textwidth}
\centering
\includegraphics[ height=5cm]{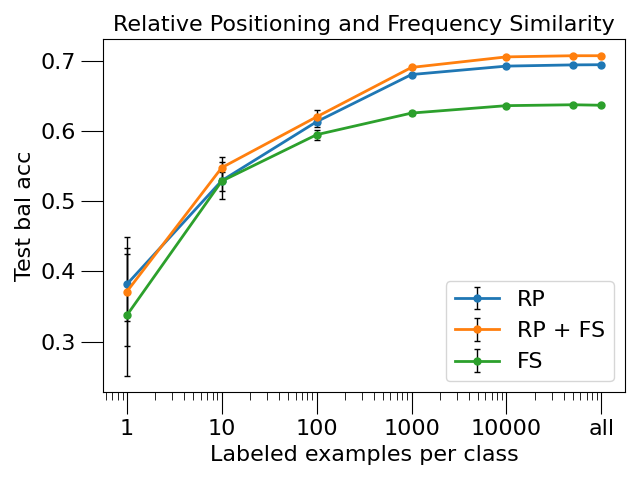} 
\caption{}
\label{fig:subim1}
\end{subfigure}
\begin{subfigure}{0.44\textwidth}
\centering
\includegraphics[height=5cm]{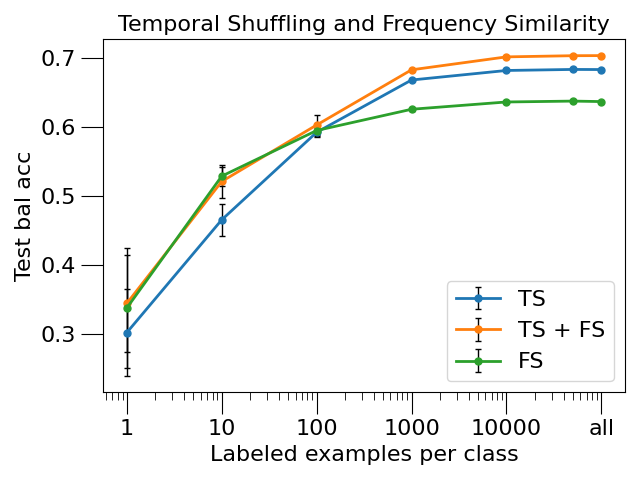}
\caption{}
\label{fig:subim2}
\end{subfigure}

\caption{(a) Test balanced accuracy of sleep stage classification with RP and FS features at different levels of labeled examples. (b) Test balanced accuracy of sleep stage classification with TS and FS features at different levels of labeled examples.}
\label{fig:image2}
\end{figure}

\begin{figure}[h!]
\includegraphics[width=0.44\textwidth, height = 4.3cm]{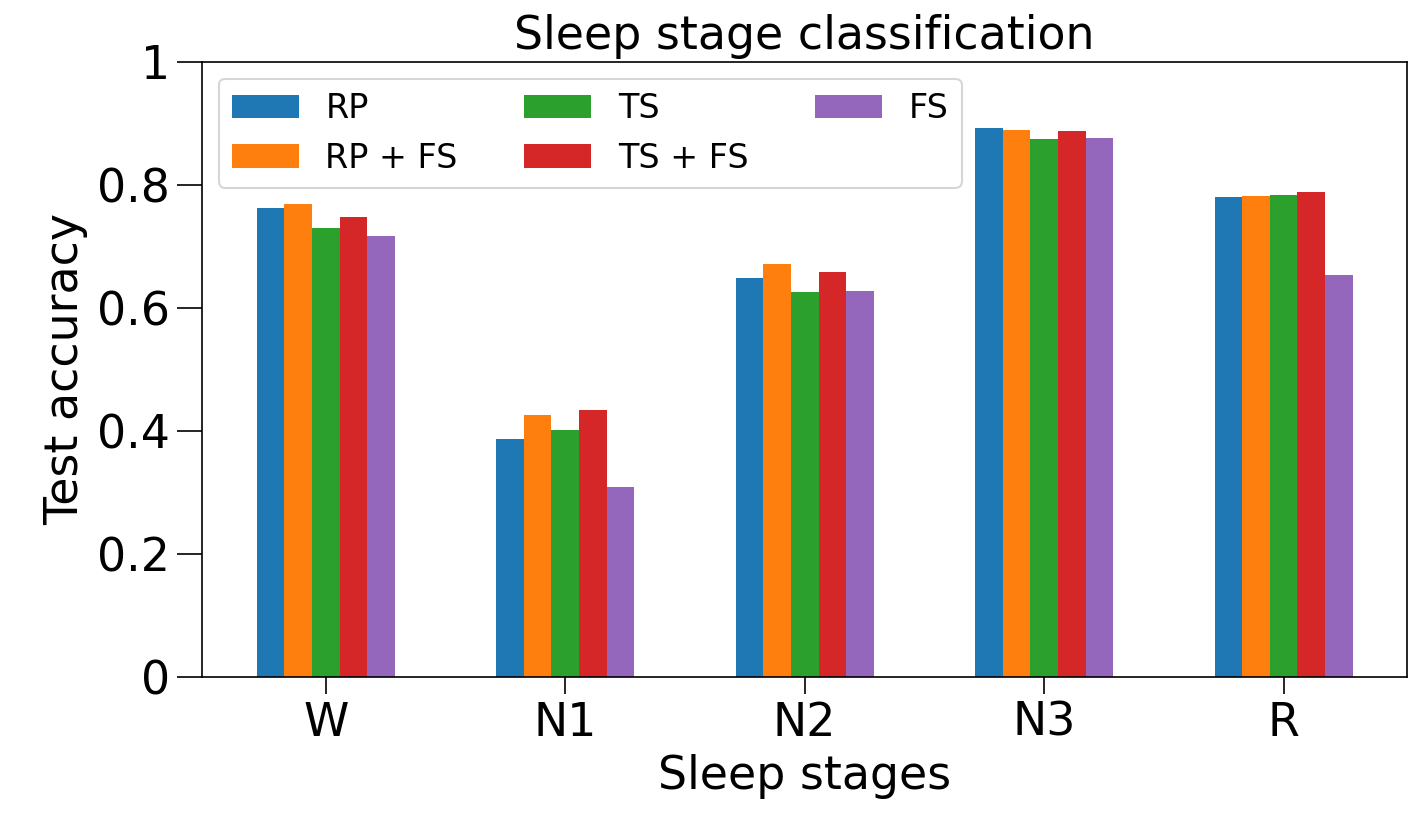}

\caption{Test accuracy of each of the different pretext tasks and their combinations over the five sleep stages, W (wakefulness), N1, N2, N3 (non-REM sleep stages) and R (REM sleep).}
\label{fig:classes2}
\end{figure}

In addition, we also examine the performance of the models by varying the number of labeled examples per class. In particular, we evaluate the models on 1, 10, 100, 1000, 10000, 50000, or all labeled examples per class and average the results over five iterations. The obtained results are depicted in Fig. \ref{fig:image2}. 
The results show that the incorporation of FS consistently improves performance across different levels of labeled examples, except when utilizing one labeled example per class. This discrepancy may be attributed to increased variability in the selection of labeled examples with a single instance, a phenomenon also observed in a previous study \cite{Banville2021}. The improvement observed across different levels of labeled examples can be attributed to FS offering valuable additional and distinct information associated with the frequency domain. Fig. \ref{fig:classes2} displays the accuracy achieved for each individual sleep stage category. Notably, when considering the integration of FS, the results show either equal or improved accuracy in comparison to the original RP and TS pretext tasks. This finding underlines the significant value added by FS as an enhancement over RP and TS. It's important to emphasize that while FS alone may not reach the same performance levels as RP and TS, it provides additional insights from the frequency domain, ultimately leading to an improvement in overall performance.

\begin{figure}[t!]

\begin{subfigure}{0.225\textwidth}
\centering
\includegraphics[height=4cm,width = 4.5 cm]{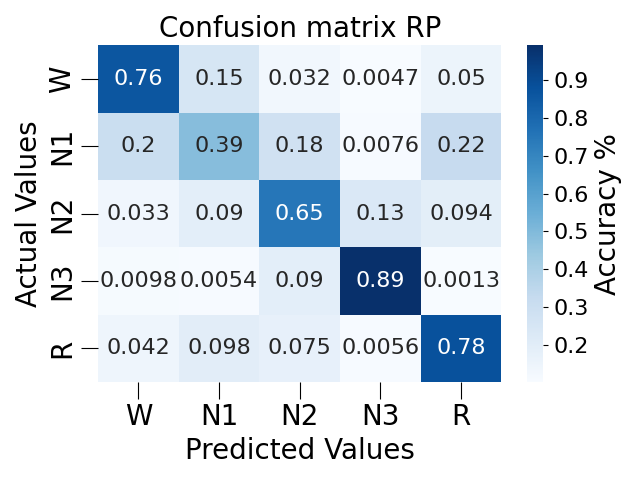} 
\caption{}
\label{fig:subconf1}
\end{subfigure}
\begin{subfigure}{0.225\textwidth}
\centering
\includegraphics[height=4cm,width = 4.5 cm]{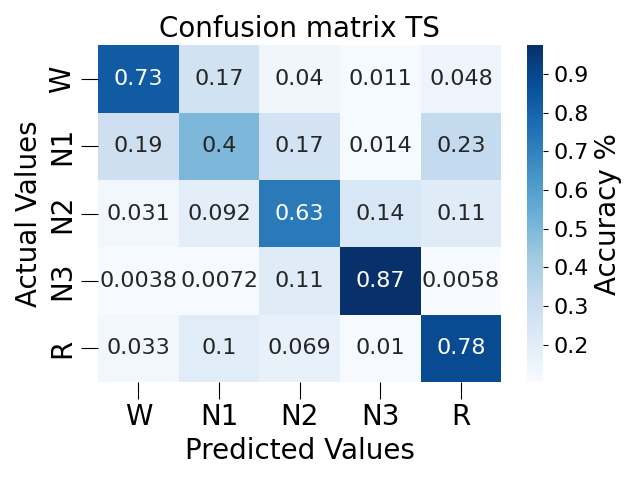}
\caption{}
\label{fig:subconf2}
\end{subfigure}

\begin{subfigure}{0.225\textwidth}
\centering
\includegraphics[ height=4cm,width = 4.5 cm]{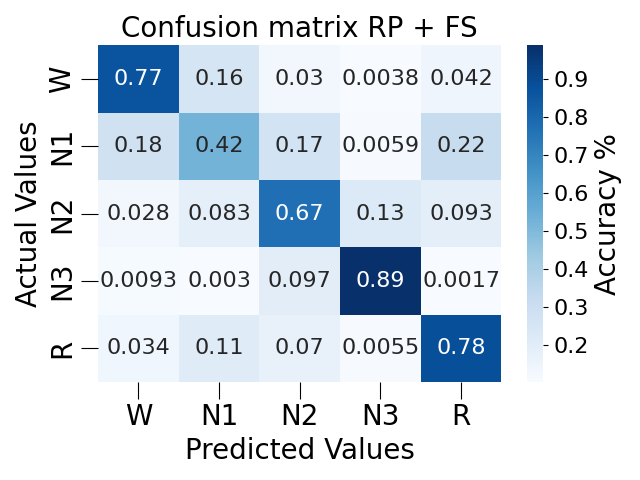} 
\caption{}
\label{fig:subconf3}
\end{subfigure}
\begin{subfigure}{0.225\textwidth}
\centering
\includegraphics[height=4cm,width = 4.5 cm]{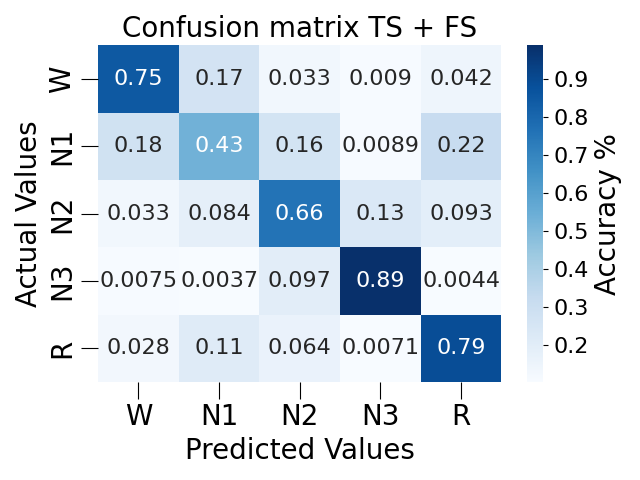} 
\caption{}
\label{fig:subconf4}
\end{subfigure}
\caption{(a) The confusion matrix of RP. (b) The confusion matrix of TS. (c) The confusion matrix of RP and FS. (d) The confusion matrix of TS and FS.}
\label{fig:confuse}
\end{figure}

Fig. \ref{fig:confuse} illustrates the confusion matrices of the pretext tasks. It can be observed that when incorporating FS into RP or TS, the accuracy remains equal or increases. This improvement primarily results from a slight enhancement in the classification of each sleep stage. Instead of solely improving the classification of one class, it contributes to a modest improvement across all classes. Furthermore, an interesting observation from Fig. \ref{fig:subconf1} is that the majority of misclassifications of the wakefulness (W) stage occur when it is predicted as N1. This could be due to the close temporal proximity between the N1 and W sleep stages.

\subsection{U-MAP visualizations of learned features}
\begin{figure}[h]

\begin{subfigure}{0.5\textwidth}
\centering
\includegraphics[width = 8.8cm, height = 0.12\paperheight]{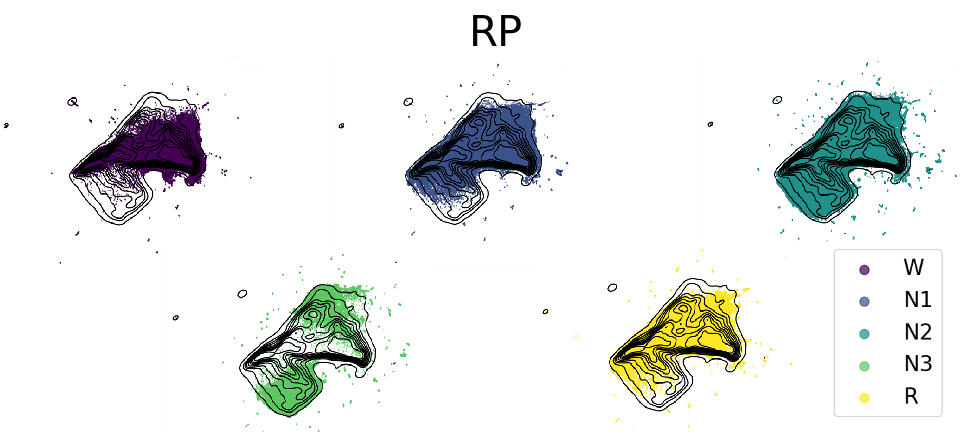} 

\caption{}

\vspace{1pc}
\end{subfigure}

\begin{subfigure}{0.5\textwidth}
\centering
\includegraphics[width = 8.8cm, height = 0.12\paperheight]{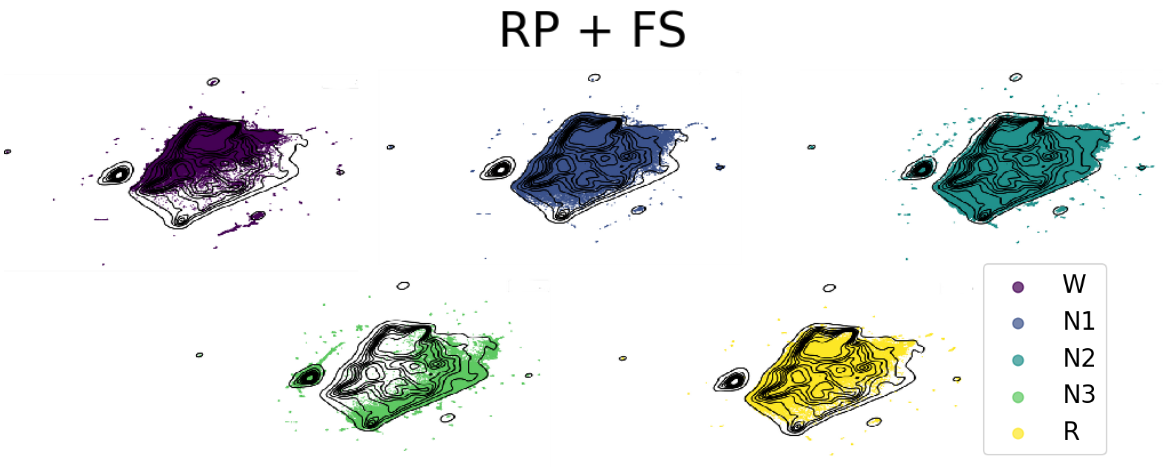} 
\caption{}

\end{subfigure}
\caption{The UMAP visualizations of the five sleep stages W (wakefulness), N1, N2, N3 (non-REM sleep stages) and R (REM sleep). (a) Relative Positioning (RP). (b) Relative Positioning + Frequency Similarity (RP + FS).}
\label{fig:umaprp}
\end{figure}

To gain deeper insights into the results and the features acquired through these pretext tasks, we utilize UMAP  (Uniform Manifold Approximation and Projection) \cite{umap2016}, a dimensionality reduction technique designed to preserve the global data structure. UMAP visualizations provide insight into the meaningful representations acquired through self-supervised learning. In Fig. \ref{fig:umaprp}, the UMAP visualizations of the features of RP and the features of RP and FS are presented. These visualizations clearly reveal the formation of distinct clusters in both the features of RP and of RP and FS, indicating that the learned features have meaningful representations of the original EEG data. However, some overlap between clusters suggests that additional information may be required for more precise sleep stage predictions, especially for distinguishing between N1 and N2 stages. The features of RP and FS offer an improvement over solely the features of RP as it enhances the separation between stages W (wakefulness) and N3. Notably, in Fig. \ref{fig:classes2}, stage N3 exhibits the highest accuracy and a similar pattern is observed in Fig. \ref{fig:umaprp} where the clearest clusters are predominantly associated with this stage.

\begin{figure}[h!]

\begin{subfigure}{0.5\textwidth}
\centering
\includegraphics[width = 8.8cm, height = 0.12\paperheight]{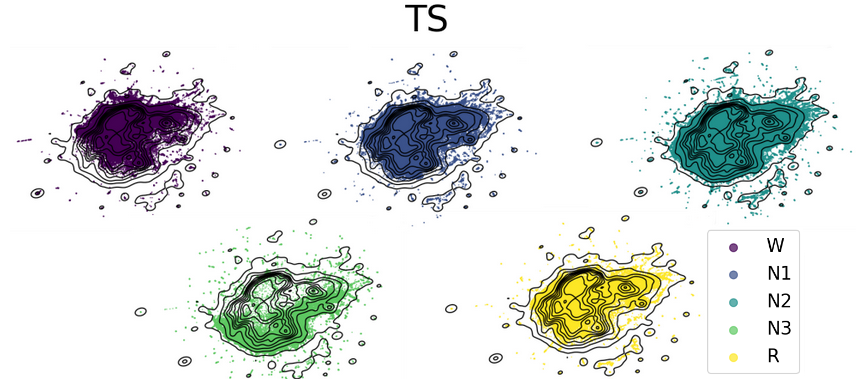} 

\caption{}

\vspace{1pc}
\end{subfigure}

\begin{subfigure}{0.5\textwidth}
\centering
\includegraphics[width = 8.8cm,height = 0.12\paperheight]{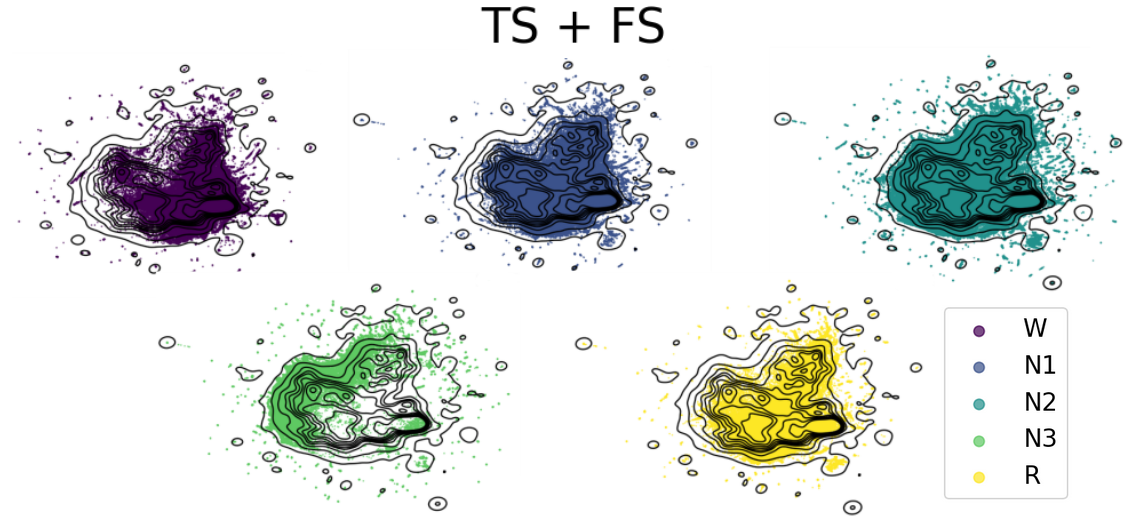} 
\caption{}
\end{subfigure}
\caption{The UMAP visualizations of the five sleep stages W (wakefulness), N1, N2, N3 (non-REM sleep stages) and R (REM sleep). (a) Temporal Shuffling (TS). (b) Temporal Shuffling + Frequency Similarity (TS + FS).}
\label{fig:umapts}
\end{figure}

 UMAP visualizations of TS and FS are depicted in Fig. \ref{fig:umapts}. Various clusters are visible, and the incorporation of Frequency Similarity features appears to improve the distinction between stages N3 and R. This observation is further supported by Fig. \ref{fig:subconf2} and Fig. \ref{fig:subconf4}, where the incorporation of FS results in reduced misclassifications between N3 and R stages.



\section{Conclusion}
In conclusion, in this paper we propose an improvement to the Relative Positioning (RP) and Temporal Shuffling (TS) pretext tasks by introducing the Frequency Similarity (FS) task, which integrates frequency information into time-based features. This augmentation leads to notable performance improvements, demonstrating the potential of self-supervised learning in sleep staging tasks. Furthermore, UMAP visualizations reveal clusters within the obtained EEG representations, confirming their meaningfulness. The proposed approach facilitates the identification of sleep disorders like sleep apnea and narcolepsy while reducing the need for costly and time-consuming manual annotations.

\bibliographystyle{ieeetr} 
\bibliography{cas-refs}

\end{document}